\documentclass[journal=jacsat,manuscript=article]{achemso}

\usepackage{graphicx}
\usepackage{dcolumn}
\usepackage{bm}

\usepackage[utf8]{inputenc}
\usepackage[T1]{fontenc}
\usepackage{mathptmx}
\usepackage{etoolbox}
\usepackage{xr}
\usepackage{hyperref}
\makeatletter
\def\@email#1#2{%
 \endgroup
 \patchcmd{\titleblock@produce}
 {\frontmatter@RRAPformat}
 {\frontmatter@RRAPformat{\produce@RRAP{*#1\href{mailto:#2}{#2}}}\frontmatter@RRAPformat}
 {}{}
}%
\makeatother

\title{\textbf{Interface Structure and Electronic Properties in Cubic Boron Nitride - Diamond Heterostructures} }
\author{Cody L. Milne}

\affiliation{Department of Physics, Arizona State University, Tempe, AZ 85287, USA}
\author{Hector Gomez}
\affiliation{Materials Science Engineering Program, University of California, Riverside, CA 92521, USA}
\author{Adway Gupta}
\affiliation{Department of Physics, Arizona State University, Tempe, AZ 85287, USA}
\author{A. Glen Birdwell}
\affiliation{RF Devices and Circuits Branch, Army Research Directorate, DEVCOM Army Research Laboratory, Adelphi, Maryland 20783, USA}
\author{Sergey Rudin}
\affiliation{RF Devices and Circuits Branch, Army Research Directorate, DEVCOM Army Research Laboratory, Adelphi, Maryland 20783, USA}
\author{Elias J. Garratt}
\affiliation{RF Devices and Circuits Branch, Army Research Directorate, DEVCOM Army Research Laboratory, Adelphi, Maryland 20783, USA}
\author{Bradford B. Pate}
\affiliation{Naval Research Laboratory, Washington, DC 20375, USA}
\author{Tony G. Ivanov}
\affiliation{RF Devices and Circuits Branch, Army Research Directorate, DEVCOM Army Research Laboratory, Adelphi, Maryland 20783, USA}
\author{Arunima K. Singh*}
\affiliation{Department of Physics, Arizona State University, Tempe, AZ 85287, USA}
\email{arunimasingh@asu.edu}
\author{Mahesh R. Neupane*}
\affiliation{Materials Science Engineering Program, University of California, Riverside, CA 92521, USA}
\affiliation{RF Devices and Circuits Branch, Army Research Directorate, DEVCOM Army Research Laboratory, Adelphi, Maryland 20783, USA}
\email{mahesh.r.neupane.civ@army.mil}
\date{\today}
\begin{document}
\begin{abstract}
Heterointerfaces of cubic boron nitride (cBN) with diamond have garnered significant interest due to their ultra-wide bandgaps and small lattice mismatch ($\sim1.5$\%), offering promising advancements in high-power and high-frequency electronic devices. However, the realization of this heterointerface has been limited by challenging growth conditions and insufficient understanding of interfacial properties. In this work, we employ density-functional theory to investigate the structural and electronic properties of diamond/cBN heterostructures as a function of interfacial stoichiometry, cBN thickness, and surface termination and passivation. Formation energies and interfacial bond lengths reveal that boron-terminated heterojunctions are the most stable while abrupt nitrogen-terminated heterojunctions are least stable, but can be stabilized by carbon-mixing. Bandstructures are computed for the heterostructures using hybrid functionals, where we find the abrupt boron-terminated and nitrogen-terminated heterojunctions exhibit $p$-type and $n$-type conductivity, respectively, while carbon-mixed heterojunctions retain wide insulating bandgaps ($4.2-4.4$ eV). The effective masses of the abrupt interfaces are found to vary strongly with stoichiometry. Intriguingly, charge analysis reveals two-dimensional electron or hole gas regions with ultra-high densities on the order of $10^{14}$ cm$^{-2}$, with distinct spatial localization on either side of the interface. Band alignments show type-I and type-II band offsets tunable by interfacial composition. Further analysis of the band alignments reveals that the diamond valence bands consistently lie above the cBN valence bands by $0.25-2.1$ eV. Interestingly, the interface termination type switches the relative conduction band position of diamond relative to the cBN conduction band, exhibiting a type-I to type-II band alignment transition. This tunability of charge dynamics and band alignment offers new design space for diamond/cBN heterojunctions. Overall, our study provides a design framework for continuously configurable electronic properties in diamond/cBN heterostructures for $n$-type field-effect transistors.
\end{abstract}
\maketitle

\section{Introduction} \label{sec:introduction}

Cubic boron nitride (cBN) is a promising ultra-wide bandgap (UWBG) semiconductor due to its array of excellent structural, electronic, and thermal properties. It shares excellent properties with other UWBG materials, such as high mechanical strength, superior thermal conductivity, and ultra-wide bandgap (6.4 eV \cite{Izyumskaya2017,Tsao2018}), which leads to high observed breakdown fields of up to 6 MV/cm\cite{Brozek1994}. However, the use of cBN films is currently limited as a result of the challenging growth of single-crystal, high-purity, and epitaxial cBN. The challenging growth of cBN is largely due to the existence of multiple BN crystalline phases, including hexagonal, cubic, and the wurtzite BN phase\cite{Milne2023,Milne2024}. Among the various substrates used to grow cBN, for e.g., Si or AlN\cite{Litvinov1998,Zhang2007}, diamond provides an intuitively suitable substrate for the growth of cBN due to their similar properties and smaller lattice mismatch. Diamond/cBN heterointerfaces have potential applications in $n$-type field effect transistors (FETs) utilizing their superb electronic properties and high thermal conductivities\cite{Shammas2017, Storm2022, Jia2023, SinghR2022,Mullen2024}. Furthermore, cBN as an intermediary layer could facilitate the integration of diamond with other III-V semiconductors when the diamond is used as a heat spreader\cite{Tsao2018}. 

There are several theoretical studies of diamond/cBN heterojunctions on the low index surfaces: the (111), (100) and the non-polar (110) surfaces\cite{Pickett1988,Lambrecht1989,Pickett1990,Yamamoto1998,Zhao2019,Jia2023,Zhu2023,Mullen2024}, guiding ongoing experimental efforts with state-of-the-art growth techniques such as electron cyclotron resonance (ECR) microwave chemical vapor deposition (MWCVD)\cite{Chan2003,ZhangW2004,ZhangW2005,Shammas2015, Vishwakarma2025}, molecular beam epitaxy (MBE)\cite{Hirama2014,Hirama2019,Storm2022}, and plasma-enhanced chemical vapor deposition (PECVD) \cite{Shammas2017,Brown2023}. However, there is a lack of fundamental understanding of the interfacial atomic structure and its role in modifying structural as well as electronic properties of the heterostructure.

The polar (100) orientation is the most common crystal orientation for heteroepitaxial growth of cBN films. This is aided by the mature growth technology for the analogous (100) diamond surface and its advantages over the (111) surface, such as more atomically flat surfaces, increased growth rates, and lower propensity for forming twins\cite{BiswasRice2025, Brown2023, Storm2022, Patel2025, Chu1992, Vishwakarma2025}. Earlier work on this surface mainly considered the superlattice approach using functionals that are known to be inaccurate in the prediction of structural and electronic properties of solids. 

More recent studies on the (100) diamond/cBN interfaces considered a limited number of interfacial configurations. In these studies, attempts were made to establish correlations between the interfacial structure and the carrier dynamics at the interface. A paper by Wu \emph{et al} studied abrupt C-B, C-N, and carbon-mixed (100) diamond/cBN interfaces and utilized accurate functionals, while the thickness dependence of the cBN slab model and band offsets for all interfacial structures were not studied\cite{Wu2020}. Many recent studies have investigated the heterostructures using state-of-the-art computational methods, but there is still a gap in the understanding of mixed interfaces, specifically the band offsets and effective masses that emerge from the complex chemistry at the interfaces. \cite{Zhao2019,Wu2020,Jia2023,Mullen2024} 

Since complex interfacial configurations are possible for polar/nonpolar interfaces in diamond/cBN heterostructures, there is a need for a more comprehensive study that includes various possible interfacial configurations that can be realized in experimental growth conditions. Furthermore, BN thickness-dependent studies could also provide insight into engineering the heterostructure properties; however, thickness-dependence has not yet been investigated from theory. In particular, studies have not yet examined the effect of the thickness of the cBN or diamond slabs in a thin-film heterointerface model along with atomistic variation at the interface. 

In this work, we study the role of interfacial compositions and cBN thickness on (100) diamond/cBN heterointerfacial stability and electronic properties by employing \emph{ab initio} first-principles calculations. For all of the heterostructures considered in this study (total 80 different configurations), we analyzed structure, thermodynamic stability and atomic bonding while providing a quantitative description of the charge transfer across the interface. Furthermore, we study the nature of their interfacial bands using hybrid functionals, their impact on the formation of two-dimensional (2D) electron and hole gas, carrier dynamics, and band alignment, and compare and contrast our results with available experimental data and theoretical results.

\section{Computational methods}\label{sec:methods}
The structural and electronic simulations in this study were performed using the Vienna \emph{Ab Initio} Simulation Package (VASP) \cite{Kresse1, Kresse2, Kresse3, Kresse4} package. The Perdew-Burke-Ernzerhof (PBE) \cite{Perdew19} exchange-correlation functional was used under the generalized gradient approximation (GGA), while the projector augmented wave (PAW) method was used for the electron-ion interaction. Structural optimization was performed for each of the heterostructures until the forces were within $10^{-2}$ eV/\AA. The cut-off energy was set at 520 eV and electronic convergence was performed for each structure until the total energy difference between steps was $<10^{-5}$ eV. 

To predict the configurations of the diamond/cBN interface likely to be observed experimentally, we evaluate their interfacial formation energies using \autoref{eq:formation_energies_cBN_diamond} after relaxation. The lattice parameters are fixed to that of diamond ($a=3.525 $\AA) and the ionic positions are then allowed to relax fully in order to simulate a diamond substrate. Vacuum regions of at least 30\AA \ are used to eliminate interactions due to periodic boundary conditions, and the dipole correction was applied to remove the spurious field in the vacuum region. The formation energies are calculated based on reference energies of the corresponding ground state elemental bulk compounds $i$ from the Materials Project database\cite{MaterialsProject}. The formation energy, $E_f$, is evaluated as:

\begin{equation}\label{eq:formation_energies_cBN_diamond}
  E_f = \frac{E_{\mathrm{tot}} - \sum_i \Delta n_{i} \mu_i
  - \sum_j n_j \mu_{j}}{A},
\end{equation}

where $E_{\mathrm{tot}}$ is the total energy of the heterostructure, $i$ is N or B, and $j$ is N, B, BN, C, and H. $\mu_{i}$ is the chemical potential of the bulk material $i$ and $n_i$ is the number of atoms of material $i$. $\Delta n_{\mathrm{B}}$ and $\Delta n_{\mathrm{N}}$ are the number of additional atoms of species B or N respectively, and $A$ is the interfacial area. The $\Delta n_{\mathrm{B(N)}}$ terms are used to correct for the non-stoichiometric nature of the cBN slab, and is defined as $\Delta n_{\mathrm{B(N)}} = n_{\mathrm{B(N)}} - \mathrm{min}(n_{\mathrm{B}}, n_{\mathrm{N}})$, while $n_{\mathrm{BN}} = 2\times\mathrm{min}(n_{\mathrm{B}}, n_{\mathrm{N}})$. Bulk elemental energies are computed with respect to diamond phase C, cubic boron nitride, diatomic N$_2$, diatomic H$_2$, and B in the $\alpha$-Boron phase.\cite{MaterialsProject} The bulk chemical potentials are the free energies computed for each bulk compound which neglect the role of vibrational entropy. This has been shown to have a small effect on the stability of functionalized diamond surfaces\cite{Gomez2024}. 

 A $\Gamma$-centered $10\times10\times1$ $k$-grid was implemented for all PBE heterostructure calculations, while bulk calculations (PBE and Heyd-Scuseria-Ernzerhof (HSE06) hybrid functional\cite{Heyd2003, Heyd2004, Heyd2006}) calculations were performed using a $\Gamma$-centered $12\times12\times12$ $k$-grid. Heterostructure bandstructure calculations are performed at 20 $k$-points per $k$-path using PBE. HSE heterostructure bandstructure calculations are performed using a $\Gamma$-centered $5\times5\times1$ $k$-grid and 6 $k$-points per $k$-path with an electronic convergence criterion of $10^{-5}$ eV. To address the bandgap underestimation problem, the HSE06 exchange-correlation functional was also used with a modified mixing ($\alpha$ = 0.33) parameter, hereafter referred to as HSE06*. This $\alpha$ was chosen to minimize the discrepancy between the theoretically predicted and experimentally measured bandgaps of diamond and cBN. With these parameters, bulk bandgaps were calculated to be 5.67 eV and 6.22 eV for diamond and \emph{c}-BN, respectively, both within 3.1\% of the experimental bandgaps of 5.5 eV and 6.4 eV bandgaps\cite{Shammas2017, Tsao2018}, respectively. Information on the hybrid functional bandgap screening is located in the Supplementary Information Table S1 and Figure S1. A comparison between bandstructures for the heterostructures calculated with PBE and with HSE06 is shown in the Supplementary Information Figure S2.

The smallest considered heterostructure models were used for HSE06* bandstructure calculation to keep the bandstructure computations tractable. We show that there is little difference in the bandstructures observed between heterostructures of different thicknesses of cBN (11 and 19 monolayers) in Figure S3 in the Supplementary Information. For calculation of the bandstructures, cBN surfaces are terminated by the dimer configuration with pseudo-hydrogen adsorbates with valencies 0.5 and 1.5 for the nitrogen and boron dimer surfaces\cite{Gong2022,Jia2023} to eliminate the metallic surface states.

In addition to the energy gaps and band edges in the bandstructure, carrier dynamics at the diamond/cBN interface will be governed by the carrier effective masses ($m*$). The carrier effective mass links the electronic structure with the carrier dynamics, and is inversely proportional to the carrier mobility. We calculate $m*$ for all interfacial compositions by fitting the interfacial band edges (calculated at the PBE level for their increased $k$-grid density) to a parabola through the relationship $E(k) = \frac{-\hbar^2 k^2}{2 m_0 m*}$, where $m_0$ is the electron mass\cite{Neupane2016}. This is performed in the directions normal to the crystal growth direction, $\Gamma - \mathrm{X}$ and $\Gamma - \mathrm{M}$.

Band offsets (BO) were obtained by using the DFT computed macroscopic-averaged electrostatic potentials (MAEP) of the heterointerface as a reference energy\cite{vandeWalle1985, vandeWalle1987, Romanyuk2013}. The valence BO (VBO) between diamond and cBN is defined by

\begin{equation} \label{eq:band_offset_C_cBN}
  \mathrm{VBO}(\mathrm{het}) = \Delta E_{\mathrm{bulk-C}}^\mathrm{VBM} - \Delta E_{\mathrm{bulk-cBN}}^\mathrm{VBM} + V_d
\end{equation}

where $\mathrm{VBO}(\mathrm{het})$ is the relative valence band offset between the diamond and cBN slabs and $\Delta E_{\mathrm{bulk-C (cBN)}}^\mathrm{VBM}$ represents the bulk valence band maximum (VBM) of C (cBN). $V_d$ is the dipole potential, which is defined as the difference between the MAEPs of C and cBN at the interface from the heterointerface calculation. In the heterostructures, the diamond substrate is fixed to its bulk lattice parameters while the cBN overlayer is then strained only in the in-plane direction, perpendicular to the growth direction $z$. Therefore, unstrained bulk diamond and bulk cBN with only in-plane strain are used to calculate the bulk band edges. In cBN, the non-isotropic strain lifts the triple-degeneracy of the VBM at $\Gamma$. Therefore, we use the average of the split valence band energy levels for $E_{\mathrm{bulk-cBN}}^\mathrm{VBM}$. 
Linear fits were performed to the average MAEP in the bulk-like regions of each C and cBN slab and extrapolated to the interfacial layer in order to calculate $V_d$, considering the non-zero electric fields within the slab regions induced by the inherent charge imbalance of the non-polar/polar interface. An example of this MAEP process is illustrated in the Supplementary Information Figures S4 and S5. This method has been successfully used to predict the VBO of similar heterojunctions between MgGeN$_2$, ZnO, and GaN alloys \cite{vonpezold2004, Kaewmeechai2020}. Potential profiles are calculated at the PBE level, since it has been shown\cite{Weston2018}, and we have confirmed, that they are nearly identical when calculated using the HSE06 functional at the same $k$-point density. Additionally, the cBN surface passivation (bulk-like, dimer, or dimer+psH) does not substantially impact the predicted band offset. Therefore, all band offsets are calculated using the heterostructures with bulk-like (monomer) cBN surface terminations. Band offsets are calculated using the PBE exchange-correlation functional for calculation of bulk band edges as in \autoref{eq:band_offset_C_cBN}. Due to the slight bandgap underestimation (overestimation) for cBN (diamond), bulk HSE06* band edges result in a constant shift of the PBE VBO by 0.42 eV upward and the PBE conduction band offset (CBO) by 0.79 eV upward in all cases.

\section{Results and Discussion}
\subsection{Structures} \label{sec:structures}

\begin{center}
\begin{figure}[t!]
 \includegraphics[width=0.5\columnwidth]{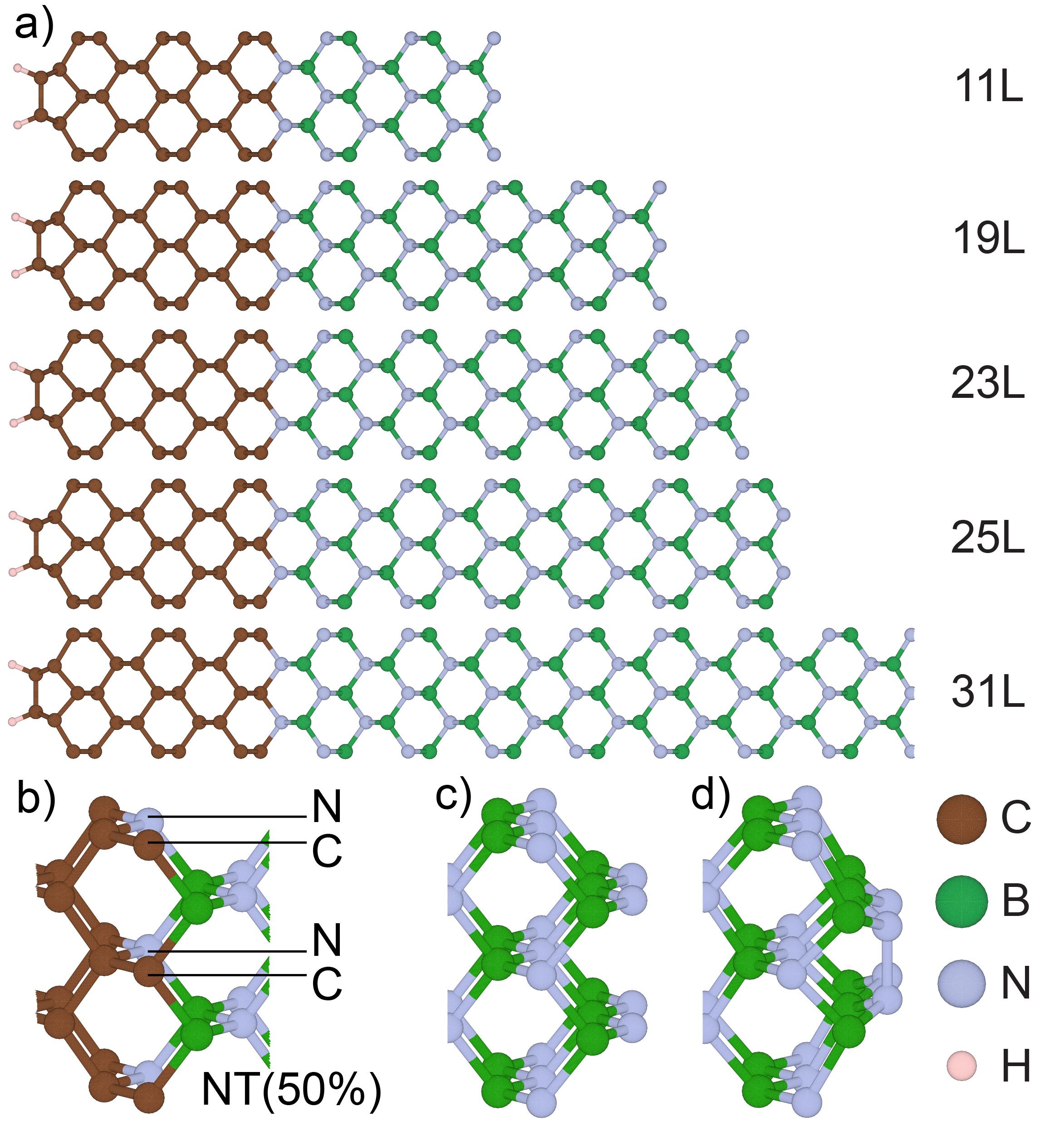}
 \caption{a) Diamond/cBN heterostructure models for the nitrogen-terminated heterostructures of 11L, 19L, 23L, 25L, and 31L cBN slab thicknesses. b) A magnified interface with carbon intermixing in the NT(50\%) heterostructure is shown. Two of the interfacial nitrogen atoms have been replaced by carbon, giving a carbon-mixing of 50\%. c) Nitrogen-terminated heterostructure with the bulk-like (monomer) surface termination. d) Nitrogen-terminated heterostructure with the dimer surface termination. Hydrogen, carbon, boron, and nitrogen atoms are shown in pink, brown, green, and gray circles, respectively.}
 \label{fig:cBN_diamond_boron_structures}
\end{figure}
\end{center}

The heterostructure models analyzed in this study consist of a 13-layer diamond (100) substrate and a cBN slab of varying thickness. The bottom dangling bonds of the diamond (100) substrate were terminated by H atoms with charge of 1. The cBN layers were varied between 11- to 31-layer slabs (corresponding to thicknesses of 9.1, 16.3, 20.0, 21.8, and 27.2\AA), with 4 atoms per layer, as shown in \autoref{fig:cBN_diamond_boron_structures}a. By default, the cBN surface is terminated by a bulk-like (monomer) (\autoref{fig:cBN_diamond_boron_structures}c) configuration. However, dimer cBN surfaces (\autoref{fig:cBN_diamond_boron_structures}d) have been shown to be thermodynamically stable\cite{Gong2022}, and are also considered in this work. Nitrogen-terminated (NT) and boron-terminated (BT) cBN models were developed for each cBN slab, which are terminated on both sides by layers of N or B, respectively. 

To elucidate the role of elemental composition at the interface, carbon intermixing was implemented into the first interfacial N- or B-layer by substituting the N or B atoms with C atoms. In total, each cBN slab thickness can have C/N or C/B mixing percentages of 0\%, 25\%, 50\%, and 75\%. The interfacial compositions are described using the convention AT(N$_\mathrm{C}$/N$_\mathrm{A}$) where A refers to the termination of the cBN slab (B or N), and N$_\mathrm{C}$ refers to the number of C atoms in the interfacial layer. The heterostructure models can be further described by their various cBN slab thicknesses and surface terminations using the convention AT(N$_\mathrm{C}$/N$_\mathrm{A}$)-M where A refers to the termination of the cBN slab (B or N), N$_\mathrm{C}$ refers to the number of C atoms in the interfacial layer, and N$_\mathrm{A}$ refers to the number of A atoms in the interfacial layers. M refers to the number of cBN monolayers in the heterostructure. If the cBN surface is terminated by the dimer configuration, "d" is attached to the end of the name. Therefore, NT(50\%)-11d refers to a heterostructure consisting of the diamond slab and 11 layers of nitrogen-terminated cBN that is terminated by the dimer configuration. At the interface, 2/4 = 50\% of the interfacial nitrogen atoms are replaced by carbon. \autoref{fig:cBN_diamond_boron_structures}b shows a visual example of interface carbon-mixing for the NT(50\%) interface. The number of cBN layers in the heterostructure is labeled NL where N is the number of cBN monolayers.

We considered 80 different heterostructure models to account for various interfacial stoichiometries, cBN thicknesses, and cBN surface terminations. The mixing percentages of BT(0\%) and NT(0\%) represent sharp C-B and C-N interfaces, which have been considered in most previous studies\cite{Zhao2019,Wu2020,Jia2023,Mullen2024}. To a lesser extent, the BT(50\%) and NT(50\%) structures have also been studied\cite{Guomin2001,Wu2020}. However, the structural and electronic properties of these structures are still poorly understood. Additionally, the interfacial stoichiometries of BT(25\%), NT(25\%), BT(75\%), and NT(75\%) have never been considered in any previous studies known to the authors. 

The diamond/cBN (Group IV/III-V) heterojunction forms a polar/nonpolar interface. Thus abrupt interfacial configurations will necessarily lead to a charged interface because of the electron surplus (IV-V) or deficiency (IV-III). Early investigations on IV/III-V Ge-GaAs interfaces have concluded that the interfaces had to contain deviations from ideal atomic arrangements, since these will result in highly charged interfaces of $\pm 1e$ per interfacial atom. These charged interfaces would be unstable in semi-infinite solids due to the buildup of electric fields. However, using thin overlayers or superlattice configurations, it may be possible to maintain the ideal interface without mixing\cite{Franciosi1993, Franciosi1996}. Therefore, we investigate both the abrupt interfaces (C-N and C-B) as well as further interfacial configurations that deviate from the abrupt interface through mixing of carbon into the first interfacial cBN layer.

\subsection{Structural properties} \label{sec:structure_properties}
\begin{center}
\begin{figure}
 \includegraphics[width=0.5\columnwidth]{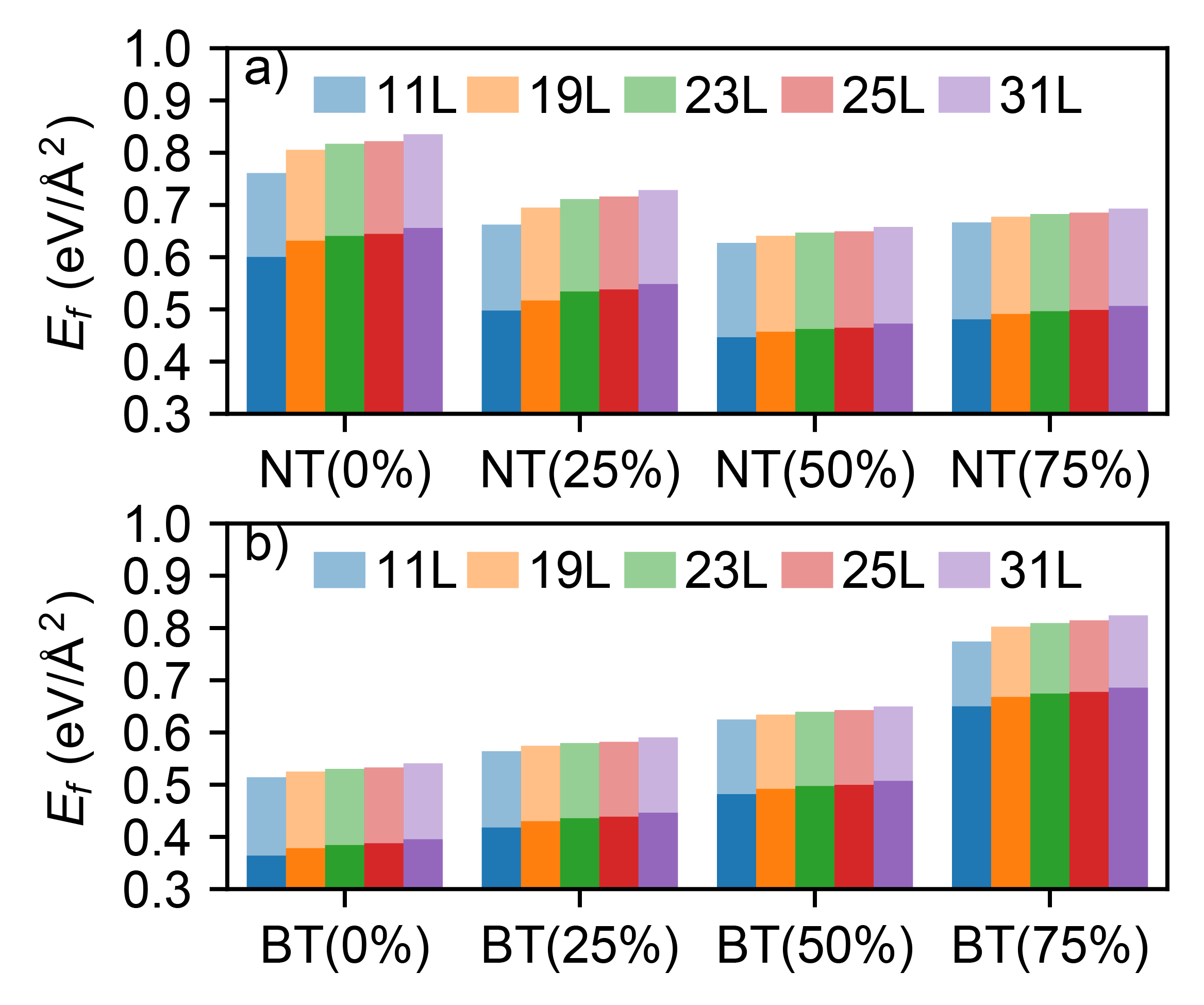}
 \caption{Formation energies of each heterostructure are shown in eV/atom, calculated with respect to each elemental bulk material as in \autoref{eq:formation_energies_cBN_diamond} as a function of interfacial carbon composition. a) Shows the formation energies of the carbon-mixed NT heterostructures while b) shows the formation energies of the carbon-mixed BT heterostructures. The formation energies of the heterostructures of 11, 19, 23, 25, and 31L cBN slab thicknesses are indicated by blue, orange, green, red, and purple colored bars, respectively. Darker colored bars show formation energies for heterostructures with dimer cBN surface reconstructions.}
 \label{fig:formation_energies_C_cBN}
\end{figure}
\end{center}

The formation energies of each simulated heterointerface are evaluated using \autoref{eq:formation_energies_cBN_diamond} and shown in \autoref{fig:formation_energies_C_cBN}. A higher formation energy demonstrates less stability. Another useful metric for interfacial stability is the work of adhesion, $W_{ad}$, which is described in Section IV and Table S2 of the Supplementary Information for the abrupt NT(0\%) and BT(0\%) interfaces.

As the thickness of the cBN slab increases, the formation energies increase in all heterostructure configurations. As the thickness of the cBN slab increases to greater than 19 ML, the increase in formation energy converges to an increase of about 0.2\% per cBN monolayer for all interfacial configurations. Therefore, this increase in energy is most likely due to the strain in cBN imposed by forced lattice matching at the interface, which raises its energy compared to that of the ground-state cBN structure. Therefore, we find that increasing the cBN slab thickness in the heterostructure requires an additional amount of energy and thus the chance of defects and strain relaxation mechanisms such as misfit dislocations induced by the lattice mismatch strain, grows in the cBN slab.

In general, interfacial stoichiometry does have an influence on the overall stability of the heterostructure. For each given thickness, the interfacial stoichiometry strongly influences the $E_f$. The BT(0\%) interface, i.e., the interface with only C-B bonding, has the lowest formation energy of all the simulated interfaces, followed by the BT(25\%) and the evenly mixed heterointerfaces BT(50\%) and NT(50\%). The similarity in formation energies between the evenly carbon-mixed (BT(50\%) and NT(50\%)) interfaces has been previously reported\cite{Guomin2001, Wu2020}, although differences in simulation methods and models make quantitative comparison difficult. This strong preference for the BT(0\%) interface is consistent with previous studies on the (100) and (111) interfaces\cite{Guomin2001,Wu2020,Jia2023,Zhu2023}. Boron has a known propensity for forming electron-deficit compounds and a strong tendency toward forming covalent bonds. As such, its bonding properties are more similar to C or Si than to the other group 13 elements\cite{Pelatt2019}. With carbon, boron forms stable boron carbide compounds such as B$_4$C\cite{Guo2021}, while carbon nitride compounds such as layered C$_3$N$_4$ are not stable under atmospheric conditions. These relationships help to explain the much lower formation energy in the BT heterostructures compared with the NT heterostructures, which are most stable in the charge-neutral NT(50\%) interfacial configuration.

Although there have been attempts to make correlations between experimental growth and interfacial stoichiometry, there has been no direct characterization of interfacial stoichiometry in most studies. Experimental techniques such as functionalization of the diamond surface prior to cBN growth could give more precise control over the interfacial stoichiometry\cite{BiswasRice2025}. Moreover, the BT interface has been directly observed in cBN grown on (111) diamond substrates\cite{Chen2015, Hirama2019}. 

To further analyze the bonding characteristics at the interface, we performed Bader charge analysis \cite{bader} to calculate the net charge transferred between the diamond and cBN slabs, shown in \autoref{tab:bader}. The charge transfer is reported here in terms of the number of electrons transferred per interfacial atom, total 8 atoms. We find that at the most stable configuration, the BT(0\%) interface, the charge transfer is maximized. Here 0.43 $e$/atom is transferred to the diamond slab. After mixing further carbon into the BT interface, the charge transfer from the cBN slab to the diamond slab decreases until at BT(75\%), the sign of the charge transfer is reversed and the cBN slab gains 0.029 $e$/atom from the diamond slab. 

On the other hand, the least energetically favorable interfaces are NT(0\%), i.e., the interface with only C-N bonding, followed by BT(75\%), and NT(25\%). We find that at the NT(0\%) interface, the cBN slab gains 0.20 $e$/atom from the diamond slab. This charge transfer decreases as carbon is mixed into the interface to a minimum of 0.021 $e$/atom at NT(25\%) mixing, and thereafter the sign of the charge transfer is reversed. At NT(50\%), the diamond slab gains 0.15 $e$/atom from the cBN slab and increases to 0.30 $e$/atom with further carbon-mixing in the NT(75\%) interface. Here we observe a relationship between the interfacial charge transfer and formation energy. The diamond/cBN heterostructure is more energetically favorable when the diamond substrate gains electrons from the cBN slab, as in the BT(0\%), BT(25\%), BT(50\%), NT(50\%), and NT(75\%) systems. However, the heterointerface is energetically unfavorable when electrons are transferred to the cBN slab, as in the NT(0\%), NT(25\%), and BT(75\%) systems. This is an indication that the polarity of the interface dictates not only the conductivity in the layer but also the overall structural stability of the heterostructure. 

\begin{center}
\begin{figure*}
 \includegraphics[width=1.0\columnwidth]{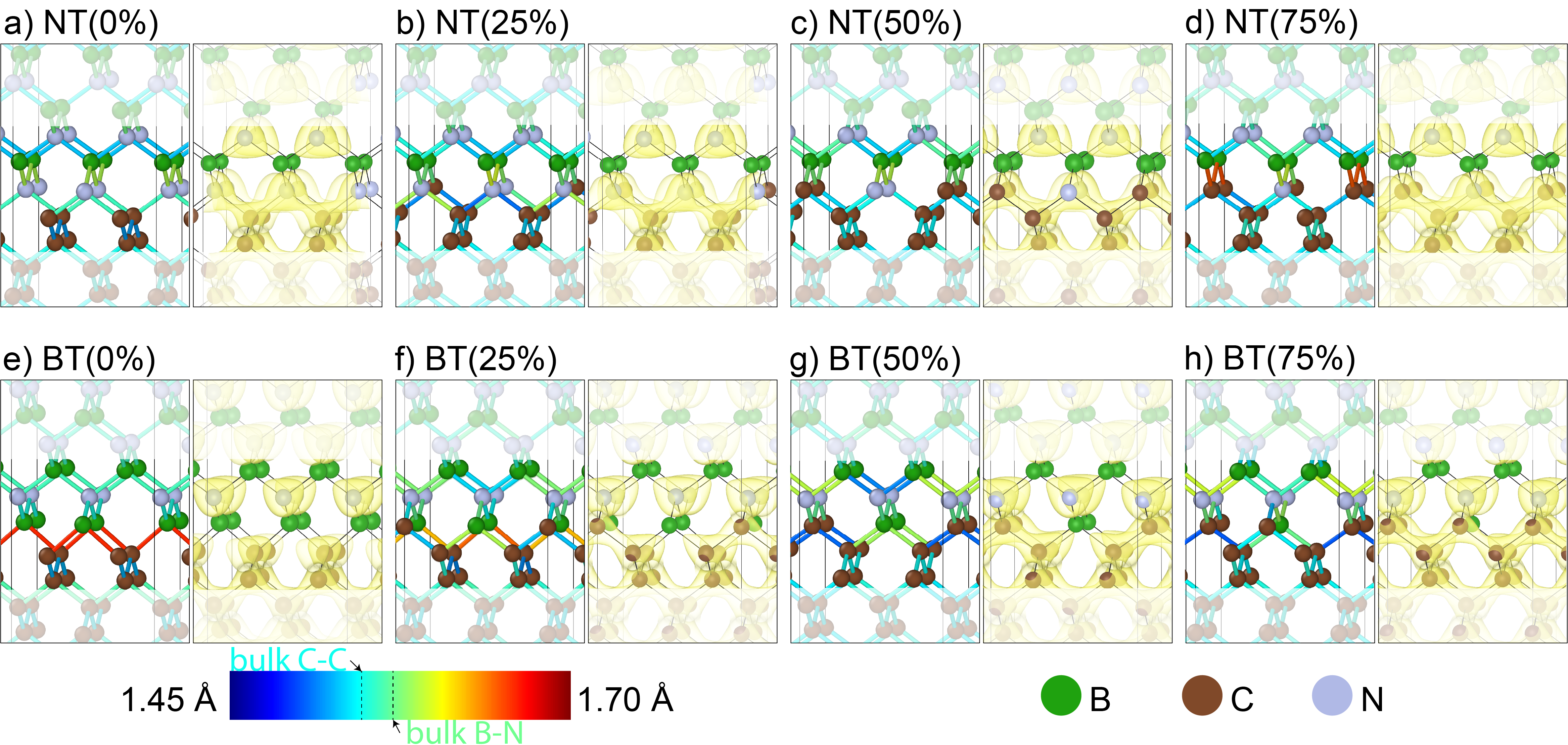}
 \caption{Strain maps and charge density isosurfaces (isosurface level 0.3 e/\AA$^3$ are shown for a) NT(0\%), b) NT(25\%), c) NT(50\%), d) NT(75\%), e) BT(0\%), f) BT(25\%), g) BT(50\%), and f) BT(75\%). Atoms and bonds far from the interface are shown as translucent for clarity. The heterostructures with 23L of cBN are used.}
 \label{fig:strain_maps}
\end{figure*}
\end{center}

The significant differences in formation energies between various heterostructure configurations can be described in part by analyzing the differences between C-B and C-N bond lengths at the interface, which can be seen in \autoref{fig:strain_maps}. The C-N (C-B) bond lengths at the abrupt interface are found to be 1.557\AA\ (1.657\AA), in line with a previous prediction of 1.547\AA\ (1.657\AA) \cite{Jia2023}. These C-B bond lengths are also similar to C-B bond lengths reported in the stable boron carbide compound B$_4$C (1.67\AA)\cite{Ivashchenko2009}. The C-N bond lengths are much larger than C-N bond lengths in graphitic carbon nitride ($d_1 = 1.345$\AA, $d_2 = 1.479$\AA) and are instead closer to that of bulk cBN or diamond\cite{Wang2018}. The C-C bond lengths nearest to the interface are found to be 1.520\AA\ (1.524\AA) for the C-N (C-B) interface, suggesting that the diamond is lightly compressively strained from its bulk value ($1.547$\AA) in both cases. The B-N bond lengths nearest to the C-N (C-B) interface are 1.585\AA\ (1.548\AA), representing tensile (compressive) strains. For comparison, our reported values for the bond lengths in bulk diamond and cBN are 1.547\AA\ and 1.570\AA\ respectively. While the C-N bond lengths are similar to those of bulk diamond and cBN, the C-B bond lengths are dramatically higher. This is shown in \autoref{fig:strain_maps}, which indicates that the C-B bond lengths are much higher than the other C-N, B-N, or C-C bonds. The C-C and B-N bond lengths nearest to the interface are similar to their bulk values, demonstrating a strain localization in the heterostructure to the interfacial layer. Moreover, the effect due to the difference in C-N and C-B bond lengths is localized to the interfacial layers and does not affect the bond lengths of the rest of the heterostructure. Naturally, incorporation of carbon-mixing into the interface will complicate the understanding of the bond lengths as the number of possible atomic bonds increases dramatically. Figure S6 in the Supplementary Information shows the average bond lengths and their standard deviation for every interfacial thickness and carbon intermixing. 

Overall, the nature of the interfacial bonding is not dramatically different from their bulk structures, as can be seen in the charge density isosurfaces in \autoref{fig:strain_maps}. As boron is bonded to both nitrogen and carbon atoms, it takes on a less cation-like character compared to B in bulk cBN. Here, more electrons are localized to the interfacial boron sites, while the nitrogen bonded to both boron and carbon atoms take on less of an anion character compared to N in bulk cBN. Fewer electrons are localized to the interfacial nitrogen sites in comparison to bulk cBN. In our study of bulk cBN, a total of 2.12 electrons are localized on the nitrogen. For comparison, in the NT(0\%) interface, only 1.35 electrons are localized on each nitrogen atom, while the interfacial carbon atoms each lose 0.38 electrons. In the BT(0\%) case, the interfacial carbon atoms each gain 0.87 electrons, while the interfacial boron atoms each lose 1.91 electrons to their C and N neighbors. After only one layer from the interface, the charge densities, Bader charges, and bond lengths do not dramatically differ from their bulk behavior, indicating a high degree of localization to the interface.

As mentioned earlier, most previous studies mainly focused on pristine interfaces, with a few studying surface reconstructions. Polar surfaces are expected to form reconstructions different from the bulk-like (monomer) configuration. The dimer reconstruction is known to provide a more stable reconstruction for the polar surfaces\cite{Karlsson2010,Jia2023}. Motivated by this, we also considered heterostructures in which the surface B or N atoms are in a dimer configuration. We have calculated the boron surface formation energies to be 0.45 and 0.29 eV/\AA$^2$ for the monomer and dimer surfaces, respectively, and 0.45 and 0.26 eV/\AA$^2$ for the nitrogen surface monomer and dimer surfaces, respectively. This dimer surface reconstruction dramatically reduces the formation energy by about 0.15 eV/\AA$^2$ in the B-surface and N-surface, as seen in \autoref{fig:formation_energies_C_cBN}. Additionally, the bottom surface of the diamond has a surface formation energy of 0.002 eV/\AA$^2$, a negligible contribution to the heterostructure formation energy. For crystal cBN growth, the formation of this reconstructed surface needs to be minimized\cite{BiswasRice2025}. The surface B or N atoms must be kept in a bulk-like configuration conducive to further cBN growth. In order to prevent this, an additional driving force in the form of higher growth temperatures and/or pressures must be supplied in the growth process. The presence of hydrogen and high bias voltages (40-100V) has been found to aid the cBN growth process, likely by helping to suppress dimerization of the surface\cite{BiswasRice2025,Brown2023}. After the growth process, it is more likely that the surface B or N atoms will arrange themselves in the dimer reconstruction, and/or will be passivated by adsorbates such as H or F atoms\cite{ZhangW2005, Shammas2015} from the growth process.

\begin{table}[]
    \centering
    \begin{tabular}{c|c|c}
        Interface & $Q$ $e$/atom & $\sigma$ ($e$/\AA$^2$)\\
        \hline
         NT(0\%) & -0.20 & -0.062\\
         NT(25\%) & -0.021 & -0.0069\\
         NT(50\%) & +0.15 & +0.047\\
         NT(75\%) & +0.30 & +0.094\\
         BT(0\%) & +0.43 & +0.13\\
         BT(25\%) & +0.30 & +0.094 \\
         BT(50\%) & +0.15 & +0.047\\
         BT(75\%) & -0.029 & -0.0094 
    \end{tabular}
    \caption{Charge transferred to the diamond slab for each interfacial configuration, where $Q$ is the number of electrons transferred to the hydrogen-terminated diamond slab from the cBN slab per interfacial atom (8). A negative number of electrons signifies that electrons are transferred to the cBN slab from the diamond slab. $\sigma$ is the total surface charge density transferred to the diamond slab per interfacial area in $e$/\AA$^2$.}
    \label{tab:bader}
\end{table}

\subsection{Electronic Properties} \label{sec:bandstructures}
\subsubsection{Bandstructures}
\begin{center}
\begin{figure*}
 \includegraphics[width=0.9015\columnwidth]{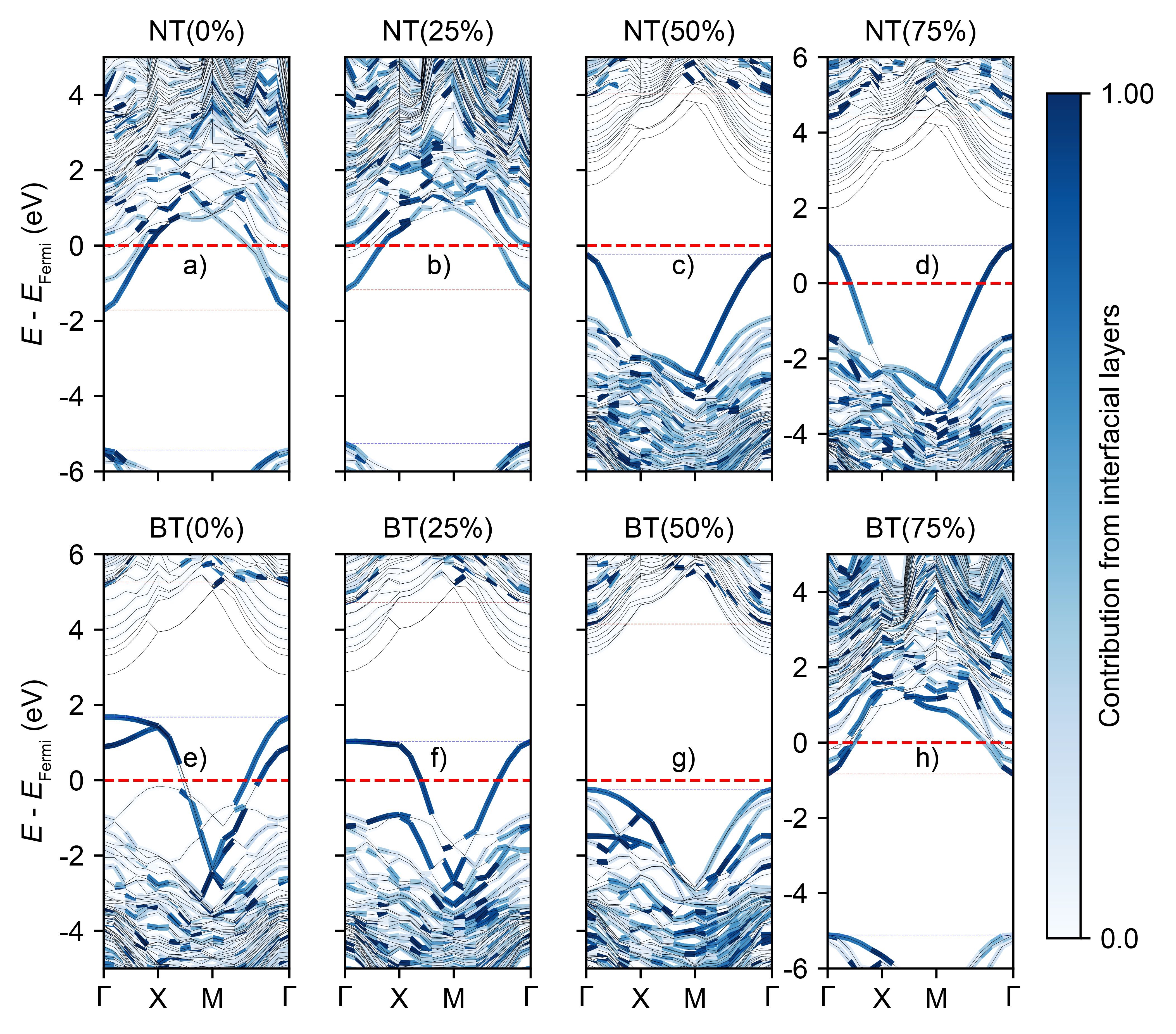}
 \caption{Electronic bandstructures in the $\Gamma - X - M - \Gamma$ $k$-path for NT and BT heterostructures. The total bandstructure of each heterostructure is shown with thin black lines while the contribution from the interfacial layers is proportional to the opacity of the blue colored lines. a) NT(0\%)-11d, b) NT(25\%)-11d, c) NT(50\%)-11d, and d) NT(75\%)-11d bandstructures are shown. e) BT(0\%)-11d, f) BT(25\%)-11d, g) BT(50\%)-11d, and h) BT(75\%)-11d bandstructures are shown. The red dashed line depicts the Fermi level for each heterostructure. Blue and green dotted lines indicate the VBM and CBM of the interfacial bands respectively. The dimer surface reconstructions are passivated with pseudo-hydrogen adsorbates with valencies 0.5 and 1.5 for the nitrogen and boron dimer surfaces respectively\cite{Gong2022,Jia2023} to eliminate the surface states. The colorbar reflects the contribution from the interfacial layers to the total bandstructure.}
 \label{fig:bandstructures}
\end{figure*}
\end{center}

Charge exchange at the interface is likely to impact the electronic properties such as the charge carrier dynamics, bandgap, and interfacial band alignment. In particular, electronic states at the interface are characterized by their topology, position relative to the Fermi level, and density\cite{Zhou2024}. In order to analyze electronic bandstructure modulation as a function of interfacial configuration, we have plotted bandstructures computed at the HSE06* level for all the configurations in \autoref{fig:bandstructures}. The bandstructures of NT(0\%), NT(25\%), NT(50\%), NT(75\%), BT(0\%), BT(25\%), BT(50\%), and BT(75\%) are shown in \autoref{fig:bandstructures}a, b, c, d, e, f, g, and h respectively. The interfacial states are colored with opacity proportional to the contribution from the interfacial layers. Aside from c) and h), all other interfaces have interfacial bands crossing the Fermi level, making them metallic. Therefore, to describe the bandgaps for each interfacial configuration, we calculate the energy gaps between the highest interfacial valence band and the lowest interfacial conduction band, labeled $E_{g, \mathrm{int}}$. These energies are shown with thin blue and red dashed lines respectively in \autoref{fig:bandstructures} and tabulated in \autoref{tab:effective_masses}. Orbital-resolved densities of states calculated at the PBE and HSE06* levels are shown in the Supplementary Information Figures S7-S13 The heterostructure bandstructures calculated at the PBE level are shown in Supplementary Information Figure S14.

In all interfacial configurations, the $E_{g, \mathrm{int}}$ are direct at $\Gamma$ and range between 3.4 and 4.4 eV, with higher $E_{g, \mathrm{int}}$ in the evenly mixed NT(50\%) and BT(50\%) interfaces and lower $E_{g, \mathrm{int}}$ in the abrupt NT(0\%) and BT(0\%) interfaces. Although the $E_{g, \mathrm{int}}$ values are still considered ultra-wide ($\geq 3.4$ eV)\cite{Tsao2018}, this is a dramatic reduction in the bandgaps in the heterostructures as compared to the bandgaps of the constituent systems, i.e., bulk diamond (5.5 eV) and bulk cBN (6.4 eV). Surprisingly, the nature of the gap is also changed to a direct gap, compared to both bulk C and cBN which exhibit indirect bandgaps. 

These differences in electronic structure due to the C-B and/or C-N bonds mark a fundamental change in interfacial chemistry, which differs from those of both bulk C and cBN. Intuitively, the interfacial mixing and nature of bonding between C and B/N could result in electronic properties similar to that of stable carbon nitride (graphitic $g$-C$_3$N$_4$) or carbon boride (B$_4$C) compounds. In fact, the reduced energy gap at the interface may be related to the fact that $g$-C$_3$N$_4$ and B$_4$C both have much smaller indirect bandgaps, approximately 2.8\cite{Ong2016, Abdullahi2016} and up to 2.1\cite{Ivashchenko2009, Gupta2022} eV, respectively. Unlike in these systems, the heterostructures presented in this study exhibit direct bandgaps. Direct bandgaps with reduced energies due to surface states have been predicted in both B- and N-terminated diamond\cite{Stacey2015, Sun2020}. Therefore, we assert that perhaps the electronic properties of the interface are more similar to B- or N-terminated diamond rather than bulk cBN or diamond. 

The reduction of bandgap at the interface is important to consider for future diamond/cBN heterojunction-based technologies, such as RF and power devices, where the magnitude of the bandgap of the channel material defines the threshold voltage and power efficiency. For example, this may result in a reduction in the critical breakdown field, since the breakdown field in a material heavily depends on the bandgap\cite{Slobodyan2022, Milne2024}. However, we observed that the carbon-mixing at the interface offers a level of tunability in the bandgap. Additionally, the directness of the bandgap means that carrier excitation between the band edges will not need to rely on phonons to impart momentum, leading to increased excited carrier concentrations for potential optical device applications.

In addition to the bandgap, the interfacial stoichiometry influences the relative position of the Fermi level. We also find that the position of the Fermi energy depends on the interfacial stoichiometry. Both $p$-type and $n$-type conductivity could be realized through control of the interfacial stoichiometry. In general, the ratio between the C-B and C-N bonds at the interface governs the overall conductivity type of the heterostructure. When the C-B:C-N bond ratio at the interface is $<1$, the system exhibits $p$-type conductivity. Conversely, when the ratio is $>1$, the system is $n$-type of nature. Importantly, the observed interfacial bond ratio-driven tunability of the conductivity could provide a route for effective $n$-type doping of diamond, for which there is a lack of suitable $n$-type dopants\cite{Tsao2018}. The charge-neutral interfaces (NT(50\%) and BT(50\%)) will have a finite bandgap, although BT(50\%) has a notably higher density of states just below the Fermi level compared to NT(50\%), which will increase the rate of charged carrier excitation and might contribute to enhanced output current during operation in RF and power devices. 

\subsubsection{Effective masses}
To describe the carrier mobilities in the heterostructures, we calculated the effective masses for the interfacial band edges in the $\Gamma-X$ and $\Gamma-M$  directions, which are tabulated in \autoref{tab:effective_masses}.

It can be seen that the nature of interfacial bands is dramatically different depending on the interfacial stoichiometry. Most notably, the high number of C-B bonds in the BT(0\%) and BT(25\%) interfaces leads to highly anisotropic valence band edges in the $\Gamma - X$ and $\Gamma-M$ directions, with huge $\Gamma-X$ effective masses. In these heterostructures, it is important to note that the interfacial bonds are oriented perpendicular to the [100] direction. However, NT interfaces show a high degree of isotropy in their interfacial valence bands and lower hole effective masses. In all cases, the lowest energy conduction bands are comparatively isotropic, and their effective masses are insensitive to the interfacial configuration. 

For comparison, we tabulated the effective masses of the band edges in bulk diamond and cBN in \autoref{tab:effective_masses}. We find that generally, the NT hole effective masses are not dramatically different than their bulk counterparts, while BT hole effective masses are much higher. The electron effective masses, although calculated from $\Gamma$ in the heterostructures due to the direct bandgap, are not substantially different than those in bulk cBN or diamond.

We also note the presence of a Dirac cone at $M$ in the BT(0\%) and BT(25\%) valence bands. This feature was previously observed in the work by Jia \emph{et al}\cite{Jia2023} and suggests the presence of quantum topological phenomena\cite{Zhou2024}. The interfacial bands due to C-N bonding display effective masses that are more similar to the bulk bands in cBN or diamond, increasing the performance of NT diamond/cBN heterostructures as $n$-type conductive layers for diamond-based electronics. However, in the BT heterostructures it is important to consider the role of the anisotropic partially occupied valence bands. The hole mobility is dramatically lower in the $\Gamma-X$ direction, with highly localized unoccupied valence band states. Density of states analysis reveals that these states can be attributed to the $p_y$ states at the interface due to the C-B bonds, mostly from the interfacial C layer, with a lesser contribution from the interfacial B layer. The large anisotropies and huge $\Gamma-X$ hole effective mass indicates that for better performance, the $\Gamma-M$ direction could be used as a hole conduction pathway instead of the $\Gamma-X$ direction in the BT heterostructure. In the following section, we investigate further the role of the partially occupied valence or conduction bands in the charged interfacial configurations.

\begin{table*}[t!]
    \centering
        \begin{tabular}{c|c|c|c|c|c|c|c|c}
        Interface &$E_{g, \mathrm{int}}$ & $m^*_{\mathrm{h}}$ ($\Gamma \rightarrow X$) &  $m^*_{\mathrm{h}}$ ($\Gamma \rightarrow M$) & $m^*_{\mathrm{e}}$ ($\Gamma \rightarrow X$) & $m^*_{\mathrm{e}}$ ($\Gamma \rightarrow M$) \\
        \hline
        NT(0\%) &  3.72$^a$ & 1.14$^b$& 0.94$^b$& 0.35$^b$& 0.34$^b$\\
        NT(25\%)  & 4.08$^a$ & 0.40$^b$& 0.67$^b$& 0.40$^b$& 0.38$^b$\\
        NT(50\%) & 4.26$^a$ & 0.29$^b$& 0.59$^b$& 0.62$^b$& 0.78$^b$\\
        NT(75\%) & 3.40$^a$ & 0.29$^b$& 0.58$^b$& 0.57$^b$& 0.63$^b$\\
        BT(0\%) & 3.59$^a$ & 8.77$^b$& 0.59$^b$& 0.65$^b$& 1.28$^b$\\
        BT(25\%)  & 3.69$^a$ & 11.11$^b$& 0.60$^b$& 0.67$^b$& 0.90$^b$\\
        BT(50\%)  & 4.39$^a$ & 3.09$^b$& 0.84$^b$& 0.70$^b$& 0.71$^b$\\
        BT(75\%)  & 4.28$^a$ & 2.95$^b$& 1.21$^b$& 0.35$^b$& 0.39$^b$\\
        \hline
        Diamond  & 5.67$^a$ & 0.26 (lh)$^a$ & 0.40 (lh)$^a$ & 1.48 (CBM $\rightarrow \Gamma/X$)$^a$ & \\
          & & 0.78 (hh)$^a$ & 1.82 (hh)$^a$ & & \\
          & & 0.29 (lh)$^b$ & 0.45 (lh)$^b$ & 1.68 (CBM $\rightarrow \Gamma/X$)$^b$ & \\
          & & 0.81 (hh)$^b$ & 2.03 (hh)$^b$ & & \\
        & & 0.29 (lh)$^c$& 0.48 (lh)$^c$ & 1.55 (CBM $\rightarrow \Gamma/X$)$^c$ & \\
        & & 0.45 (hh)$^c$ & 2.52 (hh)$^c$ & & \\
        \hline
        cBN  & 6.22$^a$ & 0.44 (lh)$^a$ & 0.55 (lh)$^a$ & 0.37 ($X\rightarrow U$)$^a$ & 0.87 ($X\rightarrow \Gamma$)$^a$\\
        & & 0.78 (hh)$^a$ & 3.40 (hh)$^a$ &  & \\
        & & 0.40 (lh)$^b$ & 0.58 (lh)$^b$ & 0.31 ($X\rightarrow U$)$^b$ & 0.95 ($X\rightarrow \Gamma$)$^b$\\
          & & 1.38 (hh)$^b$ & 3.67 (hh)$^b$ & & \\
        & & 0.45 (lh)$^c$& 0.50 (lh)$^c$ & 0.27 ($X\rightarrow U$)$^c$ & 1.15 ($X\rightarrow \Gamma$)$^c$\\
        & & 0.54 (hh)$^c$ & 3.00 (hh)$^c$ & &  \\
    \end{tabular}
    \caption{Energy gaps ($E_g$) in eV of the interfacial regions in the bandstructures and effective masses of the interfacial bands, given in $m_0$ for electrons and $-m_0$ for holes, calculated using parabolic fits to the band edges in the NT(0-75\%)-11d and BT(0-75\%)-11d heterostructures. In every case, the bandgap is a direct gap at $\Gamma$. The bandgaps and effective masses are given for bulk diamond and cBN for comparison, where lh and hh refer to light holes and heavy holes respectively.\\
    $^a$ HSE06* calculations from this work.
    $^b$ PBE calculations from this work. \\$^c$ Using $GW$ calculations from Ref. \cite{Sanders2021}. }
    \label{tab:effective_masses}
\end{table*}

\begin{center}
\begin{figure*}
 \includegraphics[width=1.0\columnwidth]{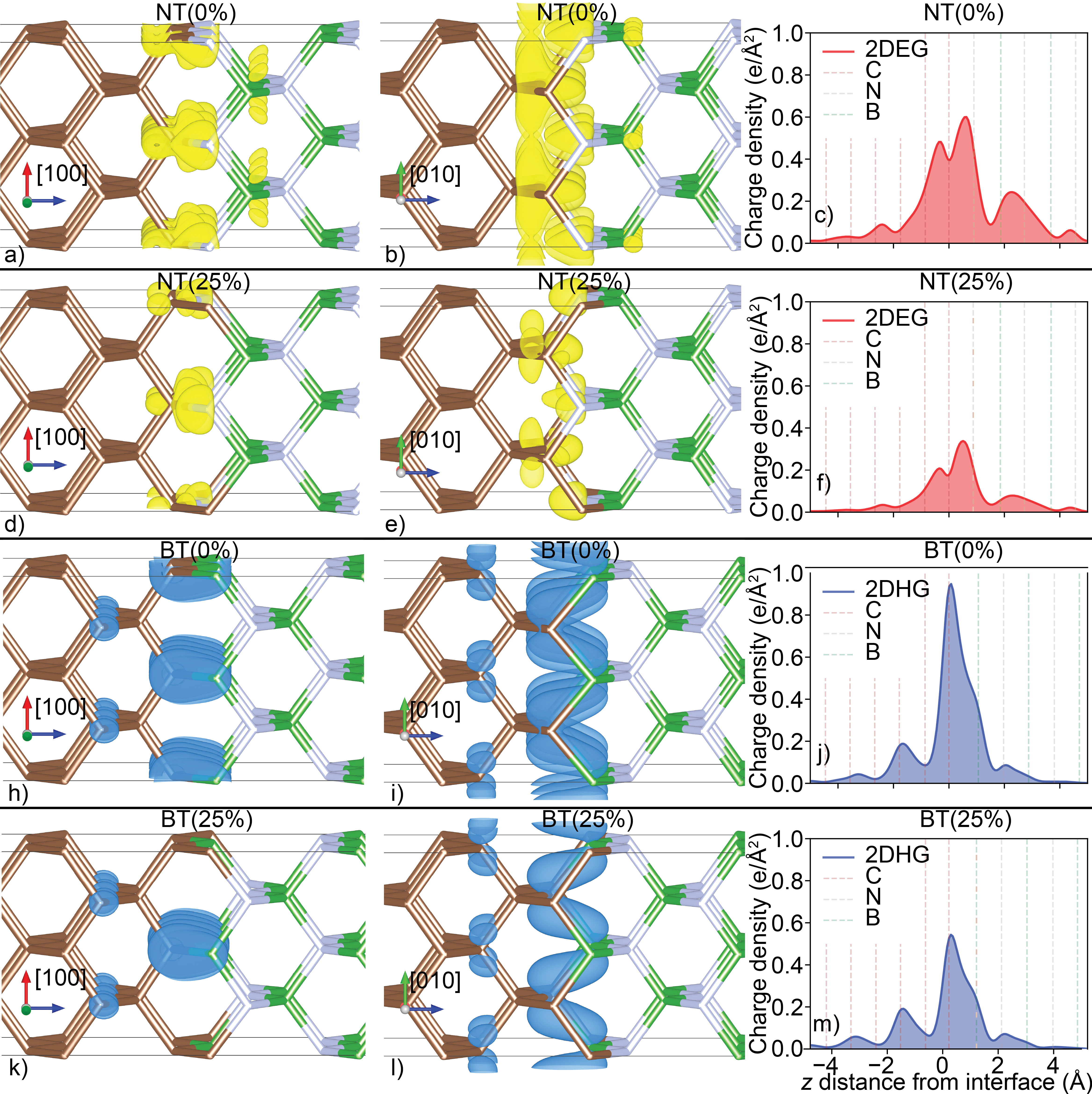}
 \caption{Shows the electron (hole) sheet charge density isosurface due to the interfacial occupied conduction (unoccupied valence) states in yellow (blue). The isosurface level is 0.004 e/\AA$^3$. The first column shows the charge density isosurface for a) NT(0\%), d) NT(25\%), h) BT(0\%), and k) BT(25\%) in the $[100]$ orientation. The second column shows the charge density isosurface for b) NT(0\%), e) NT(25\%), i) BT(0\%), and l) BT(25\%) in the $[010]$ orientation. Brown, green, and gray bars represent carbon, boron, and nitrogen atoms, respectively. The third column shows the planar-averaged charge density profiles in $e/$\AA for c) NT(0\%), f) NT(25\%), j) BT(0\%), and m) BT(25\%), where brown, green, and gray lines represent the carbon, boron, and nitrogen interfacial layers, respectively.}
 \label{fig:interface_charge_density}
\end{figure*}
\end{center}
\subsection{Interfacial charge dynamics}

At nonpolar/polar charged interfaces such as the (100) BT and NT diamond/cBN heterointerfaces studied here, the electronic reconfiguration and the presence of unoccupied valence or occupied conduction band states is expected to occur\cite{Portugal2021}. These interfacial states suggest the existence of electrons or holes spatially localized to the interface, giving rise to either a 2D electron gas (2DEG) or 2D hole gas (2DHG)\cite{Franciosi1996}. Understanding the nature and type of interfacial charges is crucial since their densities determine the switching behavior, leakage current, field-effect mobility, and threshold voltages in heterostructure-based devices\cite{Zhou2024}. In this section, we describe and quantify the localized charge carriers that are generated in many of the diamond/cBN interfacial configurations.

\autoref{fig:interface_charge_density} show the spatial distribution of the electronic-state resolved charge densities corresponding to the occupied conduction band states in the 2 eV energy region below the Fermi level in the NT(0\%) and NT(25\%) cases in yellow (\autoref{fig:interface_charge_density}a, b, d, e), and the unoccupied valence band states in the 2 eV energy region above the Fermi level in the BT(0\%) and BT(25\%) cases in blue (\autoref{fig:interface_charge_density}h, i, k, l). The right-most panels show the planar-averaged charge density profiles for the charge densities depicted in the left-most and center panels for c) NT(0\%), f) NT(25\%), j) BT(0\%), and m) BT(25\%).

In the NT(0\%) interface, the charge distribution along the $[100]$ direction (\autoref{fig:interface_charge_density}a) is fragmented in comparison with the [010] (\autoref{fig:interface_charge_density}b) orientations, indicating relatively poorer conduction along the $[100]$ orientation. This trend is also observed in the BT(0\%) interface (\autoref{fig:interface_charge_density}h and i), although the qualitative shape of the charge distribution is different. The BT(0\%) interfaces have much more charge localization in the $[100]$ orientation in comparison to the [010] orientation than in the NT(0\%) interfaces. We find that this localization is quantified in the very high $[100]$ hole effective masses discussed previously, shown in \autoref{tab:effective_masses}. Specifically, they are attributed to the interface C and B $p_y$ states. In contrast, the NT(0\%) and NT(25\%) (\autoref{fig:interface_charge_density}a, b, d, and e) charge densities are noticeably more isotropic, which is corroborated by the more isotropic band dispersion and resultant electron effective masses. These states are attributed to the $p_y$ and $p_z$ states of the interfacial layer as well as $s$ states from the interfacial N layer, with minor contributions from the $p_y$ and $p_z$ states. Hence, the NT interfaces are more spatially isotropic since the 2DEG states are from C and N $s$, $p_y$, and $p_z$ states. Contrarily, the BT interfaces are less spatially isotropic since the 2DHG states are largely attributed to only C and B $p_y$ states.

Further analysis of the planar-averaged charge density profiles (\autoref{fig:interface_charge_density}c, f, j, and m) along the $z$-direction reiterates the localized nature of these states. Most of these charges are localized within a few monolayers of the interfaces, with the highest contribution confined to the heterojunction. Qualitatively, the interfacial states are more localized to the heterojunction in the BT cases as compared to the NT cases. 

Introduction of carbon-mixing into the interface, as in NT(25\%) (\autoref{fig:interface_charge_density}d and e) and BT(25\%) (\autoref{fig:interface_charge_density}k and l), does not dramatically alter the qualitative shape of the charge density. However, it can be seen that the concentration of interfacial charge carriers is substantially reduced, leading to delocalized charges away from the interface. This is expected, since carbon-mixing in the BT or NT interfaces reduces the number of excess charge carriers at the interface by changing the C-B:C-N ratio towards 1. This also corresponds with a reduction in charge transfer between the diamond and cBN, shown in \autoref{tab:bader}.

Integration of the interfacial DOS over these states reveals the very high concentration of localized charge carriers. Quantitatively, the areal electron densities in the sharp N- and B-terminated interfaces range from $1.6 \times 10^{14}$ to $4.1\times 10^{14}$ cm$^{-2}$ respectively. These values are consistent with some previous predictions\cite{Jia2023,Ji2024}, although other studies have predicted a wider range of carrier concentrations ($10^{12}-10^{14}$ cm$^{-2}$)\cite{Narendra2019, SinghR2022, Li2024, Mullen2024}. We find that after mixing 25\% carbon into the interfaces of these heterostructures, the areal carrier densities are dramatically reduced by half to $7.8\times 10^{13}$ cm$^{-2}$ and $1.8\times 10^{14}$ cm$^{-2}$ in the NT(25\%) and BT(25\%) cases respectively. Unfortunately, comparison with experiment is unavailable since experimental studies of the diamond/cBN heterointerfacial electronic properties such as the mobility and charge density, which require high-quality interfaces, are still scarce in the literature.

Carbon-mixing of 50\% produces a charge-neutral interface, completely eliminating these states, while further mixing such as in NT(75\%) and BT(75\%) presents areal carrier densities similar to BT(25\%) and NT(25\%) respectively due to the equal numbers of C-N and C-B bonds.

Therefore, we find that the diamond/cBN interface has the potential to yield very high-density sheet charges ($\sim10^{14}$ cm$^{-2}$) of either electrons or holes at the interface. The huge concentrations of these states can be attributed to the large excess charge at the C-B (IV-III) or C-N (IV-V) interfaces and the localization of these charges near the interface. However, it is likely that impurities in real interfaces, such as carbon-mixed interfaces, could dramatically reduce the charge carrier concentrations. At these huge charge densities, interface scattering will play a major role, lowering the mobilities of the 2D carrier gases\cite{Reuters2014}. Furthermore, the interfacial charge conductivity also depends on the interfacial orientation. For example, the spatial distribution of the charge distribution and the carrier mobility are different in the NT and BT cases, due to the different electronic properties observed in the C-N and C-B bonds. The differing confinement to the interface will be strongly affected by the interface roughness in real heterostructures\cite{Reuters2014,Gurusinghe2005,Antoszewski2000}. 

The 2DEG at NT interfaces could be a novel solution for diamond $n$-type field-effect transistors (FET)\cite{Li2024}. The high theoretical maximum 2DEG density is higher than that observed at AlGaN/GaN HEMT devices ($\sim 10^{13}$ cm$^{-2}$)\cite{Bharti2025} and thus may provide comparable performance when considering limitations in real devices. Also, the 2DHG at BT interfaces could provide comparable performances in diamond-based $p$-type FETs due to the high theoretical maximum 2DHG density when compared to existing technologies such as the 2DHGs formed by H-terminated diamond surfaces ($10^{12}-10^{13}$ cm$^{-2}$)\cite{Kawarada2022}. 

\subsection{Band offsets} \label{sec:band_offsets}
Discontinuities in electronic bands are expected to occur at interface between two different materials due to their different electronic properties. These band discontinuities are described by the valence band offsets (VBO) and conduction band offsets (CBO), which influence the transport of excited charge carriers across the heterointerface. Particularly, polar heterovalent interfaces, such as the polar diamond/cBN interface, demonstrate the greatest capability for band offset tuning since various interfacial atomic configurations and orientations can result in dramatically different interfacial dipoles. These interfacial dipoles are known to have a substantial contribution to the band offsets\cite{Tersoff1984,Harrison1986,Franciosi1996}. This is in contrast to nonpolar (such as (110) diamond/cBN) or isovalent interfaces (such as Ge-Si), which have band offsets that are not significantly altered by the orientation at the interface. Therefore, it is important to investigate the interfacial configuration and cBN thickness dependencies of the band offsets in the heterojunctions studied in this work for their applicability in RF and power devices applications.

The Anderson electron affinity (EA) rule has been used historically as a first approximation to estimate the VBO between two materials using the electron affinities of their surfaces\cite{Franciosi1996}. Here, we define a positive VBO or CBO as the energy of the diamond band edge above that of the cBN. Using the reported EA of 0.65 eV for (100) bare diamond\cite{Zulkharnay2022} and the EA for (100) bare cBN which range between 1.14 and 1.49 depending on surface termination\cite{Gong2022} as well as the experimental bandgap values of 5.5 and 6.4 eV for diamond and cBN respectively, the VBO should lie in the range $1.39-1.74$ eV. This range is notably similar to DFT calculations of the nonpolar (110) diamond/cBN interface (VBO = $1.33-1.45$\cite{Pickett1988,Pickett1990,Lambrecht1989,Mullen2024}).

Surface dipoles, which determine the EA, are not necessarily the same as the interfacial dipoles\cite{Franciosi1996}, hence Anderson's rule has poor agreement with experiments for polar interfaces. Therefore, we employed the potential profile method to calculate the VBO between diamond and cBN, where we use DFT methods to calculate the average macroscopic potential in the diamond and cBN layers which is used as a reference potential. This method includes the modification of the band offsets due to the interfacial dipole.

\begin{center}
\begin{figure}[t!]
 \includegraphics[width=0.5\columnwidth]{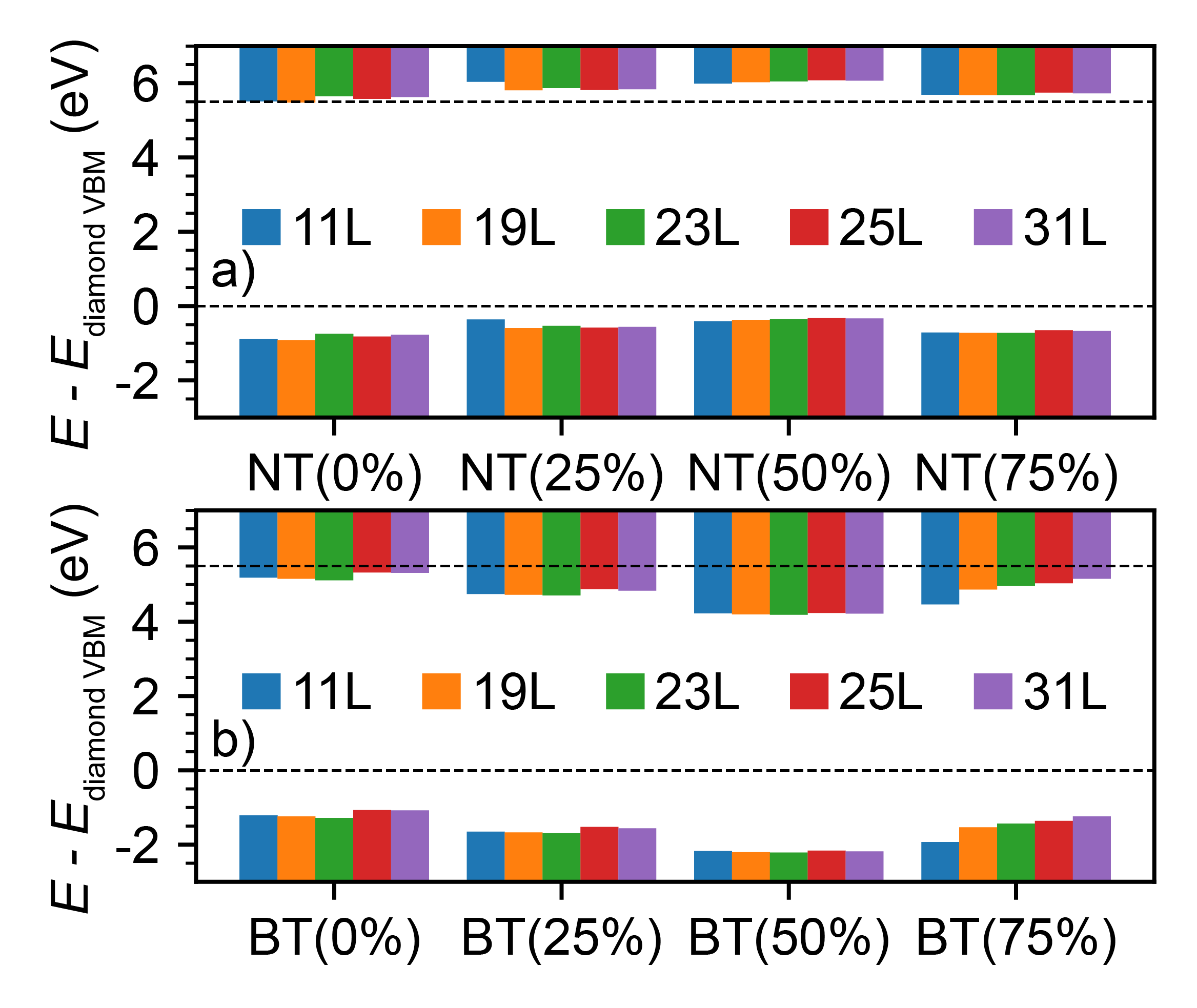}
 \caption{The band offset diagram is depicted for all interfacial compositions of the C-cBN heterostructures. a) shows the NT heterostructures while b) shows the BT heterostructures. The diamond band extrema are depicted by black dashed lines where the diamond VBM is set to 0.}
 \label{fig:band_offsets}
\end{figure}
\end{center}

The band offsets for each interfacial composition and cBN slab thickness are comprehensively shown in \autoref{fig:band_offsets} as a function of carbon-mixing at the interface for the NT (\autoref{fig:band_offsets}a) and BT (\autoref{fig:band_offsets}b) interfaces, respectively. The dashed lines indicate the band edges of diamond and the bars indicate the band edges of cBN in the heterostructure, where the VBM of diamond in the heterostructure is set to 0. Band offsets are calculated using the PBE functional with experimental bandgap values, while band offsets calculated using HSE06 with $\alpha=0.33$ are shown in the Supplementary Information Figure S15. 

As is evident from the data, the VBOs in these systems are insensitive to the layer thickness of cBN beyond 11L. In order to analyze bulk-like band alignment while varying the interface composition in these heterostructures, we utilize the system with 31L of cBN for our quantitative analysis.

Comparing the VBOs for the sharp interfaces, the VBO for the BT(0\%)-31L (1.03 eV) is higher than that of NT(0\%)-31L (0.64 eV). The electronegativity difference between B and C (0.51), is slightly higher than the difference between C and N (0.49), which perhaps contributes to the higher predicted VBO in the BT(0\%)-31L\cite{electronegativity}. Additionally, it is likely that the larger bond lengths in the BT(0\%) interfaces than the C-N bond lengths at the NT(0\%) interface, enhance the interfacial dipole strength and impact the VBO. 

The few existing DFT studies of the BT(0\%) and NT(0\%) (100) interfaces report VBO values of 0.2-0.79\cite{Wu2020,Mullen2024} and 0.34\cite{Mullen2024}, respectively. This disagreement is likely due to the use of different pseudopotentials, heterostructure treatment methods, and/or DFT calculation parameters. However, DFT studies of the BT(0\%) and NT(0\%) (111) interfaces have reported VBO values near 1.1 and 0.6 eV\cite{Zhu2023}, respectively, depending on the diamond and cBN thickness, in good agreement with our findings.

Unlike in the case of the abrupt interface heterostructures, the VBO varies with the carbon-mixing. At the 50\% C intermixing in particular, NT and BT interfaces exhibit contrasting and diverging behavior. For the NT(50\%) interface, the VBO decreases to a minimum value (0.23 eV) of all interfacial configurations. However, the value for the BT(50\%) reaches its maximum (2.12 eV). The net energy difference between the VBO values for these two cases is $\sim 1.9$ eV, which is consistent with the previous report on similar models\cite{Guomin2001}. Interestingly, this behavior, where the two inequivalent neutral (corresponding to NT(50\%) and BT(50\%) here) interfaces have the largest differences in band offsets, has been reported previously in studies of polar IV/III-V and II-VI/III-V (100) heterojunctions (Ge-GaAs and ZnSe-GaAs\cite{Franciosi1996}). While these interfacial configurations are electrically neutral, these configurations maximize the ionic dipoles at the interface, leading to their high difference. It has also been pointed out that the average of the two extreme values is remarkably close to that predicted for the nonpolar (110) interface\cite{Franciosi1996}. With our results, we find that the average of our NT(50\%) and BT(50\%) VBO values is 1.19 eV, in good agreement with VBO values of the (110) interface (VBO = $0.71-1.45$\cite{Pickett1988,Yamamoto1998,Pickett1990,Lambrecht1989,Mullen2024}). 

It has also been demonstrated that it is possible to construct mixed interfaces which are electrically neutral and the ionic dipole is eliminated\cite{Harrison1978, Franciosi1996}. These configurations require two mixed interfacial layers, similar to NT(75\%) (BT(75\%)) but with the following B (N) layer mixed with 25\% C. Interfaces with no ionic dipole should lead to offsets similar to that of the (110) interface\cite{Franciosi1996}. Our calculations confirm that the VBO predicted from these other interfacial configurations is $1.24-1.28$ eV, in excellent agreement with the average of NT(50\%) and BT(50\%).

The one existing measurement of the (100) diamond/cBN VBO is -0.8 eV\cite{Shammas2017}, in poor agreement with all prior studies. It is very likely that the presence of competing BN phases (such as h-BN) or defects related to the growth process (such as the presence of H, F, Si, O, etc.) contributes to this discrepancy. Further research is required to investigate the effects of these growth contaminants.

The predicted VBO and conduction band offset (CBO) values indicate that for the BT interfacial configurations, there is a type-II band alignment where the diamond band edges lie above those of cBN, indicating spatially separated and localized electrons and holes in the cBN and diamond layers respectively. However, the NT(0-75\%) interfaces yield CBO values which are negative ($\leq -0.26$ eV), indicating a type-I band alignment where electrons and holes are localized in the cBN layer. Furthermore, these band lineups are ideal for the use of cBN overlayers as $n$-type conductive layers. In the BT interfaces, the large VBO values ($\geq 1.0$ eV) and low to moderate CBO values ($\geq 0.13$ eV) could find potential applications for hole blocking layers, aided by their high relative thermodynamic stability. 

In summary, we observe that the (100) diamond/cBN heterojunction is demonstrated to be a remarkable system for band offset tuning by up to 1.9 eV, as well as type-II to type-I band alignment shifting, depending on careful modulation of the interfacial stoichiometry. Additionally, CBO tunability in the NT systems also offers 2DEG modulation for the realization of the diamond $n$-type FETs\cite{Li2024}.

\section{Conclusion} \label{sec:conclusion}

In conclusion, the heterointerface between cBN (100) and diamond (100) surfaces have been systematically investigated using first-principles calculations by varying interfacial stoichiometry, surface thickness, and surface passivation and termination. The overall stability of the heterostructure is governed by interfacial composition, with boron-termination preferred over nitrogen-termination. Increasing the layer thickness leads to linear increase in formation energy, which must be overcome for high-quality crystalline growth of cBN through elevated growth pressures or temperatures. Additionally, we find that the dimer surface termination in the cBN layer reduces the heterostructure formation energy by approximately $0.15$ eV/\AA$^2$ for both N- and B-terminated surfaces, and limits additional cBN growth. Interestingly, the relative stability of the heterostructures increases when the diamond substrate gains electrons from the cBN layers but decreases when electrons are transferred from the diamond layer to the cBN slab systems. Thus, we find that the polarity of the interface dictates not only the conductivity in the interface, but also the overall structural stability of the heterostructure.

Bandstructures were calculated for all heterointerfacial configurations. The abrupt interfaces, with no carbon-mixing, display $n$-type conductivity for the NT(0-25\%) and $p$-type conductivity for the BT(0-25\%) interfaces with semi-metallic behavior. However, the evenly carbon-mixing BT(50\%) and NT(50\%) heterointerfaces retain ultra-wide bandgaps ($4.2-4.4$ eV), although less when compared with bulk diamond or cBN. Though this value is $\sim25\%$ less than bulk diamond or cBN, the carbon-mixed heterostructures could still maintain large breakdown fields, similar to H-diamond\cite{Bi2020} and Ga$_2$O$_3$\cite{Kumar2025high}. Interestingly, the BT interfaces have unusually high hole effective masses with strong anisotropy, which results in very low hole mobilities. However, the NT interfaces have high electron mobilities more like the bulk compounds, which could enable excellent performance as $n$-type conductive layers. 

A detailed analysis of charge densities at the interface reveals the formation of very high 2D electron gas (2DEG) ($1.6\times 10^{14}$ cm$^{-2}$) and 2D hole gas (2DHG) ($4.1\times10^{14}$ cm$^{-2}$) concentrations at the abrupt NT(0\%) and BT(0\%) interfaces, respectively. However, the spatial distributions these interfacial charges differ between the NT and BT interfaces. This is attributed to the differences in orbitals in the C-N (C and N $s$, $p_y$, and $p_z$ orbitals) and C-B bonds (C and B $p_y$ orbitals) at the interface. Band offsets in the studied heterostructures depend strongly on the interfacial configuration as well as the carbon-mixing at the interface. The evenly carbon-mixed interfaces show the most divergent VBO values ($0.25-2.1$ eV). In the BT cases, we observe a type-II band offset; however, the NT interfacial configurations show a type-I band offset. In all cases, the diamond valence band edges lie above those of cBN by $0.25-2.1$ eV. This provides an opportunity for band alignment type switching based on modulation of the interfacial stoichiometry. 

This work expands our understanding of the diamond/cBN heterointerface and provides insight toward stabilization of the growth of cBN film with higher thicknesses and an opportunity to tune band alignment between these constituent systems. We have provided the first quantitative description of the mobility of the (100) diamond/cBN heterointerface, and studied more complex interfacial configurations (BT(25\%), BT(75\%), NT(25\%), NT(75\%)) which have not been considered in past studies. Our work demonstrates a design framework for future diamond/cBN device design, particularly using the NT interface for $n$-type conductive layers on diamond. Future work is required to further understand the complex dynamics of the diamond/cBN interface, such as the inclusion of dopants and $sp^2$ bonded BN near the interface, as well as the role of dopants and contaminants in the diamond substrate. In addition to structural stability, the composition-dependent band alignment switching in these structures provides an additional avenue for band structure engineering in these promising UWBG heterostructures.

\section*{Acknowledgments}
This work was supported by the AEOP Fellowship program with DEVCOM Army Research Laboratory and computational grants for this work were provided by the DOD High Performance Computing Modernization Program at the U.S. Air Force Research Laboratory and Supercomputing Resource Centers. This work was also supported in part by ULTRA, an Energy Frontier Research Center funded by the U.S. Department of Energy (DOE), Office of Science, Basic Energy Sciences (BES), under Award No. DESC0021230. This research used computing resources from the San Diego Supercomputer Center under the NSF-XSEDE and NSF-ACCESS Award No. DMR150006, and the Research Computing facility at Arizona State University. A.S. was also funded through a U.S. Department of Defense HBCU/MI Summer Faculty Research Fellowship.

\section*{Data Availability Statement}
Data is available upon request from the authors.

\bibliography{main}

\end{document}